\documentclass[twocolumn,showkeys,preprintnumbers,amsmath,amssymb]{revtex4}
\usepackage{graphicx,color}
\begin{document}
\title{$T_c$-Enhanced Codoping Method for GaAs-based Dilute Magnetic Semiconductors}

\author{Van An~\textsc{DINH}$^{1,2}$} 
\email{divan@cmp.sanken.osaka-u.ac.jp.}
\author{Kazunori ~\textsc{Sato}$^3$}
\author{Hiroshi~\textsc{Katayama-Yoshida}$^{1,2}$}
\affiliation{$^1$Department of Condensed Matter Physics, and {$^2$}Department of Computational Nanomaterials Design, Nanoscience and Nanotechnology Center, The Institute of Scientific and Industrial Research, Osaka University, 8-1 Mihogaoka , Ibaraki, Osaka 567-0047, Japan.\\
$^3$ Institut f\"ur Festk\"orperforchung,~Forschungszentrum J\"ulich, D-52425 J\"ulich, Germany}
\begin{abstract}
 Based on {\it ab initio} calculations of Ga$_{1-x}$Mn$_x$N$_y$As$_{1-y}$ and Ga$_{1-x}$Mn$_x$C$_y$As$_{1-y}$, we propose a new codoping method to enhance the Curie temperature $T_c$ of diluted magnetic semiconductors. The solubility of Mn can be increased  up to high concentration by the codoping of  N or C to reduce the lattice and volume expansion caused by Mn doping.  It is found that the impurity band of the majority spin is strongly broadened and pushed up into the higher energy region due to the strong $p$-$d$ hybridization caused by the codoping, and the $T_c$ becomes higher than the room temperature at $x>6\%$.
 \end{abstract}
%

\maketitle
\sloppy
%
The discovery of ferromagnetic diluted magnetic semiconductors\cite{Matsukura,ohno1,ohno2} (DMS) by incorporating the $3d$ transition metal atom (TM) into the III-V compound semiconductors was a great step towards the application of the spin degree of freedom of the electrons in semiconductor spintronics.   One of the most important issue regarding the realization of the semiconductor spintronics is to synthesize the DMS with a Curie temperature ($T_c$) higher than the room temperature. The  III-V compound-semiconductor-based DMS such as (Ga,TM)As is one of the most promising candidates from the view-point of the industrial application. Recently, the {\it ab initio} calculations\cite{Sato1,Sato3} have predicted the high $T_c$ in Mn or Cr-doped GaAs. However, up to now, the  $T_c$ which was reported in the new experiment\cite{Ku} of (Ga,Mn)As is limitted to 150K because of the low solubility of TM atoms in GaAs. 

 The $T_c$ of the ferromagnetic (Ga,Mn)As is expected to increase with both the Mn and hole concentrations\cite{ohno1,Sato1,Sato3}. In fact, the Mn atom at the substitutional site of the GaAs acts as an acceptor, but in (Ga,Mn)As, the hole concentration is substantitially lower than the Mn concentration. It is ascribable that when the TM atoms such as Mn, Cr, etc. are doped into GaAs, the lattice constant and volume of the (Ga,TM)As become larger than that of the host material, leading to the presence of compensating donors  formed by interstitial TM atoms. As pointed out by Yu {\it et al.} \cite{Yu}, the Mn atom can occupy the tetrahedral interstitial sites in the zinc-blende structure and acts as a double donor. The compensation of the substitutional Mn acceptors by the interstitial Mn donors decreases the hole concentrations and $T_c$. We can reduce the acceptor compensations due to the interstitial Mn donors by reducing the lattice or volume expansion using the codoping method.  Nitrogen (N) or Carbon (C) is one of the excellent candidates for this purpose by using the codoping of Mn and N (or C) since the N and C have the smaller atom radius than the As atom.
 
 Since N $2p$-level is lower than the As $4p$-level, we can expect strong $p$-$d$ hybridization in Mn-N than Mn-As. Therefore, the antibonding Mn $3d$-impurity state ($t_a$) is pushed up into the higher energy region in the band gap. Based on the double-exchanged mechanism of the ferromagnetism in DMS, $T_c$ is propotional to $\sqrt{x}$ and $n_{3d}(E_F)$ \cite{Sato3,Sato4,Sato5}, where $x$ is the concentration of $3d$-TM, and $n_{3d}(E_F)$ is the $3d$-TM partial density of states (PDOS) at the Fermi level $E_F$. 
 It is well known that the band structure of GaAs changes strongly with the incorporation of  isoelectronic N at As sites to form GaNAs\cite{Thordson}. At N concentration of 10\%, the GaNAs is expected to become semi-metal\cite{Thordson}. Therefore, besides the avoidance of the substitutional Mn acceptor compensation by the interstitial Mn donors, an influence of strong $p$-$d$ hybridization caused by N codoping on the magnetic properties of (Ga,TM)(N,As) to increase the $n_{3d}(E_F)$ and $T_c$ is also expected. 
 
 Following the above-mentioned idea of materials design,  we propose a new codoping method to enhance the $T_c$ of GaAs-based DMS, by which TM and N (or C) are codoped simultanously into Ga and As sites of GaAs, respectively. Incorporated concentrations of N (or C) in the codoping method are chosen to satisfy the condition of no lattice expansion or no volume expansion.

When GaAs is doped by Mn atoms, the lattice constant $a$ of the host material changes, and the dependence of the deviation of the lattice constant on the Mn concentration obeys the Vegard law\cite{ohno1}. Moreover, the volume of the Mn-doped GaAs becomes also larger. On the contrary, when some As atoms of the GaAs are substituted by N (or C) atoms, the lattice constant and volume decrease. In order to reduce the Mn donors at tetrahedral interstitial sites and increase the solubility of substitutional Mn acceptors, we set the N concentration $y$ so that the lattice constant $a$ and volume $V$ are kept unchanged based on the experimentally observed Vegard law. In other words, the dependence of $y$ on $x$ is chosen so that the deviation of the lattice constant $\Delta a(x,y)$  or the volume expansion of the lattice cells $\Delta V(x,y)$ can be ignorable. As reported by Fan {\it et al.}\cite{Fan}, the change of the lattice constant in the case of N doping with low concentration($y\le 3\%$) still obeys the Vegard law. Therefore, we can assume here that the deviation of the lattice constant is linear with N concentration. The lattice constants of 5.6533~\AA, 4.52~\AA\ and 5.98~\AA\ for GaAs, GaN and MnAs, and the atom radii of 1.221~\AA, 1.367~\AA, 1.245~\AA\ and 0.549~\AA\  for Ga, Mn, As and  N, respectively, are used to derive the dependences of the N fraction on the Mn concentration. We obtained the following relations for the options of the codoping method based upon the experiments \cite{ohno1,Fan}:$y=0.29x$ (for $\Delta a=0$), $y=0.42x$ (for $\Delta V=0$).

The calculations  are based on the KKR-CPA-LDA method \cite{KKR}. The  MACHIKANEYAMA-2000 package coded by H. Akai (Osaka University) \cite{Akai1} is used for the calculation of the density of states (DOS) and the $T_c$ of  Ga$_{1-x}$Mn$_x$N$_{y}$As$_{1-y}$. 
 In the present method, the Mn atoms replace randomly the Ga atoms in GaAs host material. The As atoms are substituted randomly by the N atoms. The $T_c$ is evaluated by using the mean field approximation\cite{Schiliemann,Sato3,Sato4}. For comparison, the calculation of the $T_c$ of the (Ga,Mn)As is also performed. 
 To simulate the spin-glass state of Ga$_{1-x}$Mn$_x$N$_y$As$_{1-y}$, the Ga cations are substituted randomly by Mn$^\uparrow$$_{x/2}$ and Mn$^\downarrow$$_{x/2}$ magnetic ions, where $\uparrow$ and $\downarrow$ denote the direction of the local moment of Mn atoms. Hence, the ferromagnetic  and spin-glass states can be decribed as (Ga$_{1-x}$,Mn$^\uparrow$$_{x}$)(N$_y,$As$_{1-y}$) and (Ga$_{1-x}$,Mn$^\uparrow$$_{x/2}$,Mn$^\downarrow$$_{x/2}$)(N$_y,$As$_{1-y}$), respectively. Throughout the calculations, the potential form is restricted to the muffin-tin type, and the muffin-tin radii of 1.224 \AA\ are chosen for both cations and anions. The wave functions in each muffin-tin sphere are expanded with real harmonic up to $l=2$, where $l$ is the angular momentum at each site. The relativistic effect is also taken into account by using scalar relativistic approximation. 328 $k$-points in the irreducible part of the first Brilouin zone are used in our calculations.
\begin{figure}
\begin{center}\leavevmode
\includegraphics[width=1.0\linewidth]{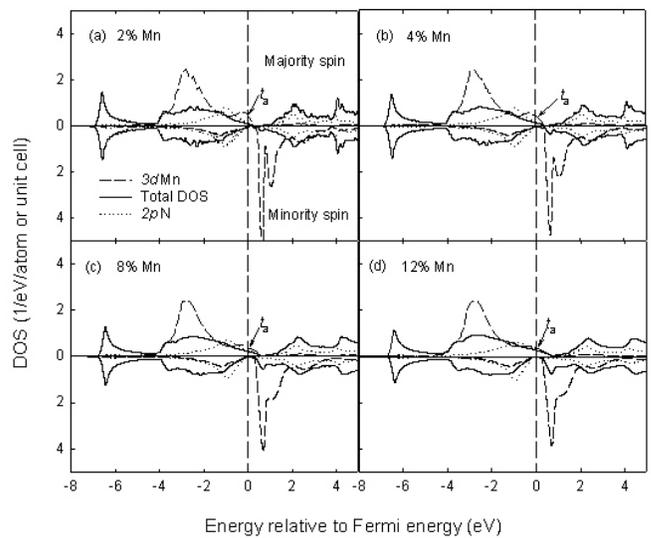} 
\caption{ Total DOS per unit cell (solid curve), PDOS of $3d$-states per Mn atom at Ga site (dashed curve) and $2p$-state per N atom at As site (dotted curve) for several Mn concentrations. N concentration is set to be 29\% of Mn concentration.}\label{f1}
\end{center}
\end{figure}

 Figure~\ref{f1} depicts the DOS of Ga$_{1-x}$Mn$_x$N$_y$As$_{1-y}$ for several concentrations of Mn ($x=2\%, 4\%, 8\%, 12\%)$. The anti-bonding states $t_a$ of $3d$-Mn caused by the hybridization with $4p$-states of As ions and $2p$-states of N ions form an impurity band in the band gap. With increseasing Mn concentration, the N concentration is also accordingly increased upon the codoping, this anti-bonding impurity band $t_a$ connects with the host valence band to widen the band width ($W$) which is proportional to the root square of the Mn concentration ($W\propto \sqrt{x}$), and the band gap is considerably narrowed (see Fig.~\ref{f1}(d)). Such a behavior relates to the well-known band gap bowing effect in the GaAs incorporated with N \cite{Weyer}.  Moreover, the Fermi level of the majority spins shiftes into the new part of the valence band to realize a half-metallic ferromagnetism. 
 
\begin{figure}
\includegraphics[width=1.0\linewidth]{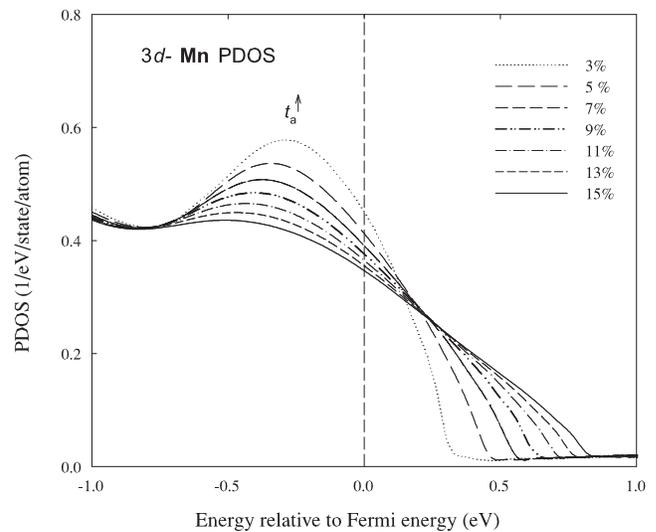}
\caption{$t_a^\uparrow$-PDOS of Mn $3d$-states in (Ga$_{1-x}$,Mn$^\uparrow$$_{x}$)(N$_y,$As$_{1-y}$). N is incorporated at As site with concentration being 29\% of Mn concentration. }
\label{f2}\end{figure}

 It is well known  that the stabilization of the ferromagnetism in GaAs-based DMS is due to the double-exchange mechanism \cite{Akai1,Sato1,Sato3}. In the DMS with the double-exchange mechanism being dominant, the $T_c$ is propotional to the band width ($T_c\propto W\propto\sqrt{x}$) and the PDOS of the anti-bonding states $t_a$ at the Fermi level ($T_c\propto n_{3d}(E_F)$). In the present case, the $t_a$-peaks are sensitive with the change of Mn concentration. The impurity band width $W$ is proportional to the Mn concentration: $W\propto\sqrt{x}$, which is shown in Fig.~\ref{f1}, and also in Fig.~\ref{f2}. 

\begin{figure} 
 \includegraphics[width=1.0\linewidth]{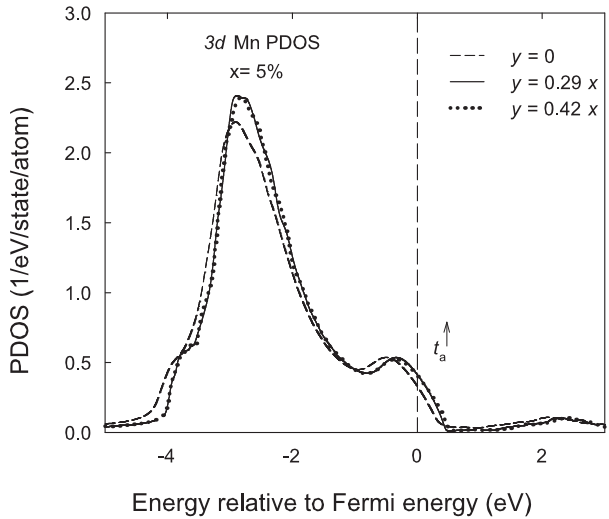}  
\caption{PDOS of Mn $3d$-states for $x= 5\%$. Incorporated N concentration is chosen as 0\% (dashed curve), 29\% (solid curve) and 42\% (dotted curve) of Mn concentration.}
\label{f3} \end{figure}

The dependence of the $t_a$-band width  of majority spins  on the Mn concentration varying from 3\% to 15\% is shown in Fig.~\ref{f3}. Similar to the Fig.~\ref{f1}, the N concentration is chosen to be equal to 29\% of Mn concentration to compensate the lattice expansion upon the codoping. It is easy to see that $t_a$-bands are gradually broadened with the substitutional Mn concentration. Since the stabilization energy of the ferromagnetism is proportional to the $t_a$-band width, it suggests to open up a realistic possibility to synthezise a high-$T_c$ DMS based on GaAs host material by the codoping of Mn and N.

For comparison, we ploted also the PDOS of the $3d$-Mn states of the ferromagnetic state (Ga$_{1-x}$,Mn$^\uparrow$$_{x}$)(N$_y,$As$_{1-y}$) at 5\% of Mn concentration of the both codoping and single-doping cases(Fig.~\ref{f3}). The dashed curve refers to the $3d$-PDOS of Mn ions in single-doped GaMnAs (i.e. $y=0$). The remainders illustrate the $3d$-PDOS for two options of the N codoping method. It is remarkable that at 5\% of Mn concentration, the $3d$-PDOSs of Mn atoms in the codoping cases show a large difference from that in the single-doping case. Because of the host valence band originates from anion $p$-states, the doping of N with $2p$-level deeper than the shallow $4p$-level of As into the host GaAs leads to the change of the energy levels of the anti-bonding $t_a$-states  with strong $p$-$d$ hybridization, and $t_a$-states become more $3d$-character. By the codoping of Mn and N, the peaks of $3d$-Mn PDOS of $t_a$-states are pushed up and closer to the Fermi level than that in the single-doping case.  The $t_a$-peaks are pushed into the higher energy region in the band gap, and the ferromagnetism caused by the  double exchange interaction is more stablized due to the higher density of $3d$-Mn PDOS at $E_F$. 

It is emphasized here that the codoped N plays an important  role to increase the $T_c$ in Ga$_{1-x}$Mn$_x$N$_y$As$_{1-y}$: 
({\it i}) Smaller atomic radius of N atoms compensates the lattice expansion caused by Mn doping at the Ga sites, then increases the solubility of Mn atoms in GaAs. 
 ({\it ii}) It changes the band structure of the host material, resulting in the shift of the  $3d$- states of $t_a$ towards the higher energy region in the band gap, the increase of the PDOS of $t_a$-states at the Fermi level, and therefore the increase of the $T_c$ caused by the enhancement of the ferromagnetic double exchange interaction.
\begin{figure}  
\includegraphics[width=1.0\linewidth]{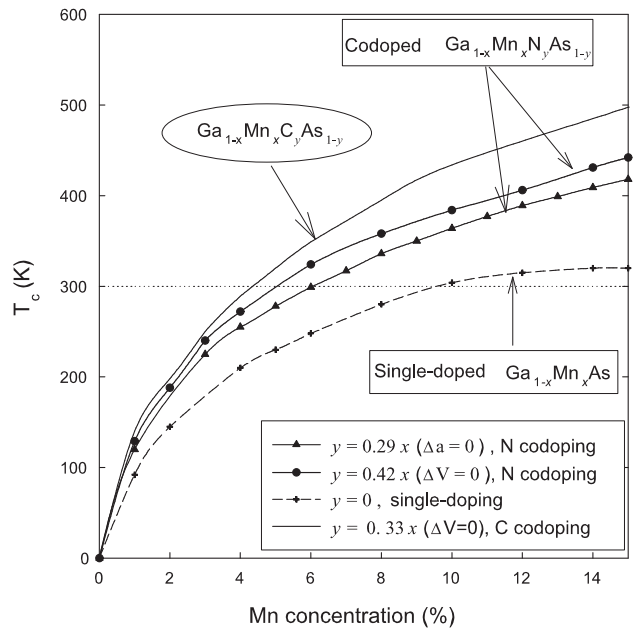} 
\caption{Curie temperature vs. Mn concentration of Ga$_{1-x}$Mn$_x$As (crosshair-dashed curve), Ga$_{1-x}$Mn$_x$N$_y$As$_{1-y}$ and Ga$_{1-x}$Mn$_x$C$_y$As$_{1-y}$ (solid curve). $y$ varies as  $0.29x$ (triangle-solid curve), $0.42x$ (circle-solid curve) for the codoping of N, and $0.33x$ for the codoping of C. The dotted line indicates the room temperature (300 K)}
\label{f4}
 \end{figure}

As usual, the estimation of the $T_c$ is performed by using a mapping on the Heisenberg model in the mean field approximation\cite{Sato3,Sato4}: $k_BT_c=2\Delta E/3x$, where $\Delta E$ is the total energy difference between the spin-glass and ferromagnetic states. Fig.~\ref{f4} shows the comparison of the $T_c$ between the single-doped Ga$_{1-x}$Mn$_x$As, and codoped  Ga$_{1-x}$Mn$_x$N$_y$As$_{1-y}$. As seen from Fig.~\ref{f4}, the $T_c$ increases with subtitutional concentration as $\sqrt{x}$ due to the ferromagnetic double-exchange interaction. The $T_c$ of the single-doped GaMnAs gains the room temperatures at 10\% of Mn concentration and reaches the saturation regime at $x~\approx~12\%$. However, the $T_c$ of Ga$_{1-x}$Mn$_x$N$_y$As$_{1-y}$ gains the room temperature at the smaller Mn concentrations ($x\ge 5\%$ for the doping option $y=0.42x$ and $x\ge 6\%$ for $y=0.29x$), and has a tendency to increase ($\sim 30\%$) with the whole of Mn concentrations.  At the higher Mn concentrations, the codoping method gradually enhances the $T_c$  due to the presence of N atoms.

It is noted that while the highest $T_c$ of Ga$_{1-x}$Mn$_x$As was reported about 110~K by Matsukura {\it et al.}\cite{Matsukura} and newly 150~K by Ku {\it et al.}\cite{Ku}, the {\it ab initio} calculations\cite{Sato1,Sato3,Sato4} predicted the much higher $T_c$ . This  discrepancy is caused by the compensation effect in the experiment since the thermal equilibrium solubility of Mn atoms in GaAs is very small ($\sim10^{16}$~cm$^{-3}$). This effect can be reduced by the codoping method as proposed in this letter, by which we can reduce the interstitial Mn atoms by N codoping with reducing the lattice expansion caused by Mn doping. However, it is noted here that since the influence of the band gap bowing effect due to N doping, the band gap becomes narrower, resulting in the $t_a$-states could be located in the valence band, then the $p$-$d$ exchange interaction becomes dominant at the sufficiently high concentration of the N atoms. Therefore, the codoping with N concentration higher than 10\% may causse the reduction of the $T_c$.

Since the $2/3$ of the $t_a^\uparrow$-bands are occupied by electron in (Ga$_{1-x}$,Mn$^\uparrow$$_{x}$)(N$_y,$As$_{1-y}$) as shown in Fig.~2, we can further increase the $T_c$ by the $p$-type codoping which increases $n_{3d}(E_F)$ and the stabilization energy from the ferromagnetic double-exchange interaction. Carbon is suitable candidate for this purpose. Using the ion radii of  0.62~\AA, 0.67~\AA, 0.58~\AA, and 0.16~\AA\ for Ga, Mn, As and  C, respectively, we derived the relation of Mn and C concentrations as $y=0.33x$ regarding no volume expansion. Our {\it ab initio} calculations show the more gradual enhancement of the $T_c$ compared with the codoping of N as demonstrated in Fig.~4.
  
  In conclusion,  we propose a new method of codoping to synthezise the High-$T_c$  DMS based on the {\it ab initio} calculations of the electronic structure and Curie temperatures of the Ga$_{1-x}$Mn$_x$N$_y$As$_{1-y}$ and Ga$_{1-x}$Mn$_x$C$_y$As$_{1-y}$. Mn and N (or C) are incorporated into host materials on condition that they do not change the lattice constant or volume of the host materials to increase the solubility of substitutional Mn atoms, as well as, the PDOS of the $t_a^\uparrow$ states at the Fermi level.

   The obtained results for Ga$_{1-x}$Mn$_x$N$_y$As$_{1-y}$ and Ga$_{1-x}$Mn$_x$C$_y$As$_{1-y}$ show the gradual enhancement of the $T_c$ by the codoping method. It is suggested that since we can reduce the compensating donors formed by the interstitial Mn atoms by reducing the lattice expansions upon the codoping, $T_c$  could be higher than the room temperature at $x>6\%$ for the codoping of N, and $x>5\%$ for the codoping of C.  The influence of the N atoms on the magnetic properties with deeper $2p-$levels of N and strong $p$-$d$ hybridization between Mn $3d$- and N $2p$-levels is also strongly expected to increase the $T_c$. Also, the $p$-type codoping of C is the promising method to gradualy enhance the $T_c$ based on the ferromagnetic double-exchange interactions.
  
  \acknowledgements{
  This research is supported by 21$^{st}$ Century - COE program from the Ministry of Education, Cultural, Sport, Science and Technology. We thank to Prof. H. Akai for providing useful KKR-CPA packge (MACHIKANEYAMA-2000).}
  

\end{document}